# Nanogranular MgB$_2$ thin films on SiC buffered Si substrates prepared by in-situ method

Š. Chromik[1], J. Huran[1], V. Štrbík[1], M. Španková[1], I. Vávra[1], W. Bohne[2], J. Röhrich[2], E. Strub[2], P. Kováč[3], S. Stanček[3]

[1]*Institute of Electrical Engineering, SAS, Dúbravská cesta 9, 841 04 Bratislava Slovak Republic*
[2]*Hahn-Meitner-Institut, Glienicker Str. 100, D-14109 Berlin, Germany*
[3]*Department of Nuclear Physics and Technology, Faculty of Electrical Engineering and Information Technology, Slovak University of Technology, Ilkovičova 3, 812 19 Bratislava, Slovak Republic*

MgB$_2$ thin films were deposited on SiC buffered Si substrates by sequential electron beam evaporation of B-Mg bilayer followed by in-situ annealing. The application of a SiC buffer layer enables the maximum annealing temperature of 830ºC. The Transmission Electron Microscopy analysis confirms the growth of a nanogranular MgB$_2$ film and the presence of a Mg$_2$Si compound at the surface of the film. The 150-200 nm thick films show a maximum zero resistance critical temperature T$_{C0}$ above 37 K and a critical current density J$_C \sim 10^6$ A/cm$^2$ at 11 K.

## 1 Introduction

Recently we have used NbN buffered Si(100) substrates for the preparation of the MgB$_2$ films grown by sequential evaporation of B and Mg with improved superconducting and transport properties [1]. In spite of this fact, the existence of a superconducting NbN buffer layer (T$_C \sim$ 10 K) is not always convenient for some applications. This is a challenge to look for another non-superconducting type of buffer layer on top of the "technological" Si substrate. The potential buffer layer can be chosen on the base of substrates which recently were successfully used for the preparation of MgB$_2$ films as sapphire [2,3], MgO [4,5], SrTiO$_3$ [6,7], SiC [8], and SiN [9]. Most of these substrates are monocrystals, often rather expensive (SiC, SrTiO$_3$). So far, the best results have been obtained by Zeng et al. [8] where high quality MgB$_2$ films were prepared using a hybrid physical-chemical vapor deposition (HPCVD) technique on top of a monocrystalline 6H-SiC substrate. To the best of our knowledge, no publication reports on the deposition of MgB$_2$ on SiC buffered Si substrate. This is a reason why we decided to apply an amorphous SiC as buffer layer on top of a Si substrate beside the fact that SiC is an interesting doping material for the powder in tube processed MgB$_2$ tapes.

The scope of this article is the preparation and the study of some structural and electrical properties of MgB$_2$ films on SiC buffered Si substrates.

## 2 Sample preparation

Amorphous silicon carbide buffer layers were grown by plasma enhanced chemical vapor deposition (PE CVD) technique described in [10]. All films were prepared on weakly doped n-type Si substrates Si(100) and Si(111) (resistivity 2-7Ωcm). Prior to the deposition, a standard cleaning was performed to remove impurities from the silicon surface, and 5% hydrofluoric acid was used to remove the native oxide from the wafer surface. The wafer was then rinsed in deionized water and dried in a nitrogen ambient. The films were deposited in a high frequency parallel-plate plasma reactor in which the frequency, the RF power, and the substrate temperature were maintained at 13.56 MHz, 0.06 Wcm$^{-2}$, and 350°C, respectively. Samples with different amounts of N were achieved by small addition of ammonia (NH$_3$) into the gas mixture of silane SiH$_4$ and methane CH$_4$ which were directly introduced into the reaction chamber [10].

MgB$_2$ films were prepared similarly to [1] by sequential electron beam evaporation of the Mg-B bilayer followed by in-situ annealing. The magnesium and boron layers were deposited on the SiC buffered Si substrates at room temperature. The deposition chamber was evacuated to a pressure of 10$^{-4}$ Pa. An excess of about 100% of Mg compared to the stoichiometric composition has been used. The amount of boron corresponds to 200 nm of a stoichiometric MgB$_2$ film. As-deposited films were in-situ heated to 280°C for 30 min in an argon atmosphere at a pressure of 0.06 Pa. Subsequently, the pressure of Ar was increased up to 16 Pa and the temperature of the heater was increased to the maximum temperature of 830°C and kept there for 10 min. Finally, the Ar pressure was raised to 10$^3$ Pa and the samples were cooled down to room temperature.

### 3. Characterization an discussion

The final MgB$_2$ films obtained on the SiC/Si substrate were 150-200 nm thick and often completely or partially covered by the remainder of the unreacted Mg. X-ray diffraction patterns were taken by using a Bragg-Brentano diffractometer. No peaks belonging to the MgB$_2$ nor to the SiC phase could be detected in any of the samples. This result implies a nanogranular character of the prepared films. This fact is confirmed by Transmission Electron Microscopy (TEM) (JEOL1200EX). The microstructure and the Selected Area Diffraction (SAD) patterns were examined at the surface of the film, at the center of the film and close to the interface with the SiC buffer layer. A conventional Ar ion milling process was used to etch the MgB$_2$ film. SAD pattern taken at the surface of the film (Fig.1a) reveals the presence of Mg$_2$Si besides diffraction rings originating from the MgB$_2$ hexagonal phase.

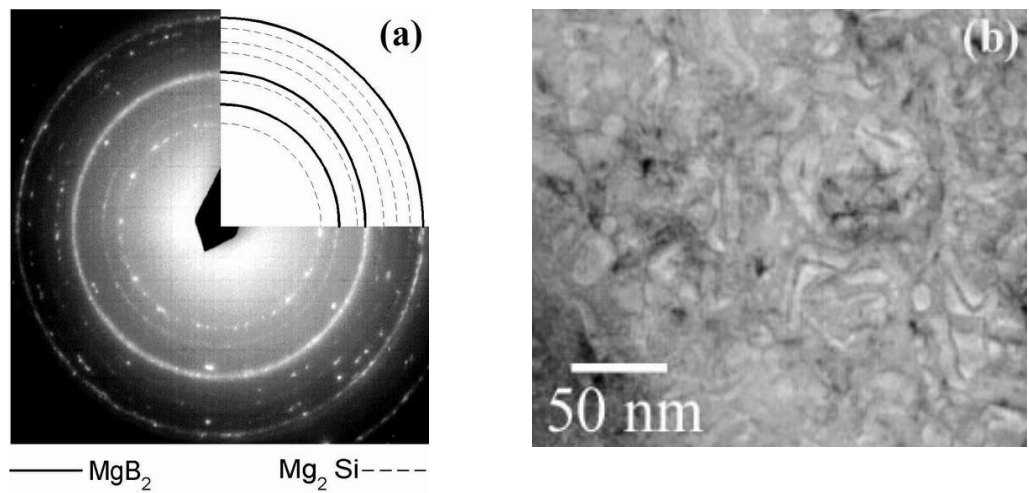

Fig.1. SAD pattern (a) and the microstructure (b) of the MgB$_2$ film surface.

A TEM micrograph (Fig.1b) depicts the presence of the $Mg_2Si$ grains with a size of about $10\times40 nm^2$. However, SAD pattern taken from the center of the film (Fig.2a) and from the region near the interface with the buffer layer shows diffraction rings belonging to the $MgB_2$, only. After the final etching, only amorphous SiC is detected near the interface (Fig.2b).

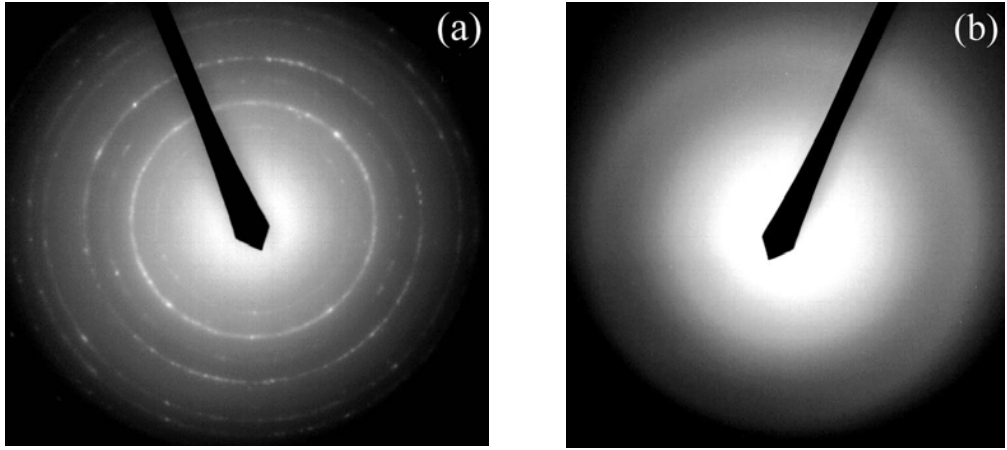

*Fig.2. SAD patterns taken from the middle of the $MgB_2$ film (a) and near the MgB/SiC interface after the final etching (b).*

The $MgB_2$ films on the SiC/Si substrates were investigated by Rutherford backscattering spectrometry (RBS). The experiments were performed with 1.7 MeV $He^{++}$ ions at a scattering angle of 168°. A RBS random spectrum similar to those obtained by Zeng et al. [8] is shown in Fig.3. The experimental points above the channel number 800 unrelated to the sample composition are a consequence of Ag paste used to fix the sample to the holder. Probably due to the instability of the beam for few seconds, this hit the edge of the film.

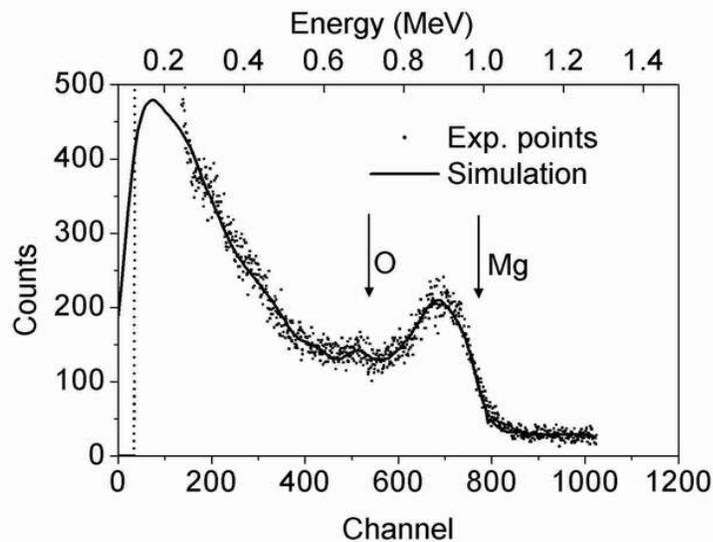

*Fig.3. RBS spectrum of a $MgB_2$ film grown on a SiC/Si substrate*

It is possible to correlate a simulated Mg depth profile with the spectrum; however, the simulation of depth profiles for B, C, and Si is difficult because the scattering signals of these elements are superimposed. The simulation of the experimental RBS spectrum shows the presence of oxygen at the surface of the $MgB_2$ film, which is a hint for a possible presence of MgO as a consequence of the oxidation of the residual Mg or $MgB_2$ surface.

The superconducting and the transport properties of the $MgB_2$ films were characterized by resistance measurements using the standard DC four-point method and by measuring of the AC susceptibility. The maximum zero resistance critical temperature $T_{C0}$ was 37.4 K (Fig.4a) and the critical transport current density $J_C$ reached about $10^6$ A/cm$^2$ at a temperature of 11 K. In Fig.4b the susceptibility measurements of this sample are shown. The diamagnetic onset was observed at a temperature of T = 37.2 K in the real part ($\chi'$) of the susceptibility and is corresponding well to $T_{C0}$. The width of the diamagnetic transition (10-90%) is 1.2 K, indicating a homogeneous $MgB_2$ superconducting phase. The imaginary part ($\chi''$) characterizes the midpoint of the diamagnetic transition (T = 36 K) and the AC losses due to this transition. The AC susceptibility measurements were made at a field amplitude of $H_0$ = 200 A/m.

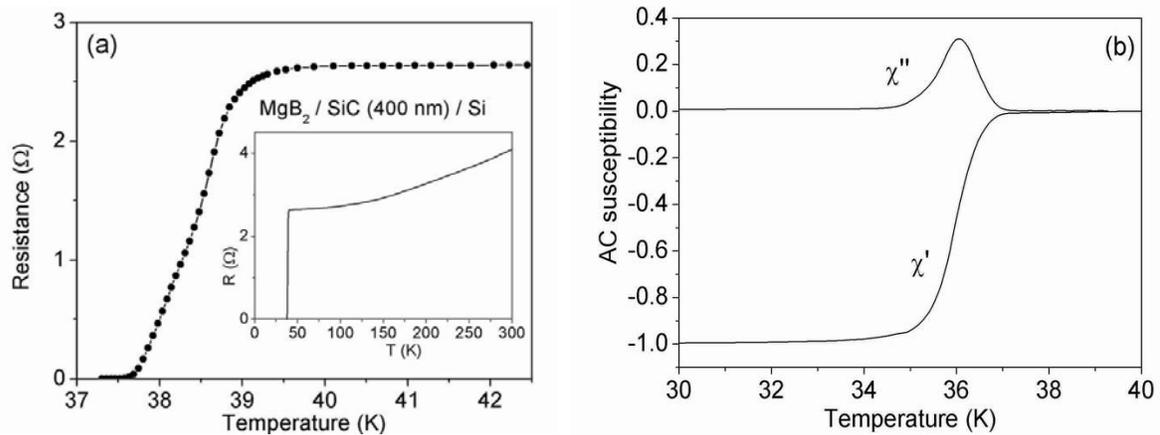

Fig.4. The dependence of the resistance on the temperature of a $MgB_2$ film (a). The susceptibility measurements of the same film (b).

The philosophy of our $MgB_2$ growth is based on the fact that the boron film is completely covered by Mg in the as deposited bilayer. We expect an increased pressure of Mg at the B-Mg interface during in-situ annealing and, therefore, an increase of the temperature up to 800 - 900°C is necessary. This temperature range corresponds to a "window" in the $P_{Mg}$-T phase diagram for $MgB_2$+Mg-gas, in which $MgB_2$ does not decompose and an excess of Mg does not condense at the $MgB_2$ surface [11, 12].

The amorphous SiC layer works as a diffusion barrier between the Si substrate and the $MgB_2$. We can expect two processes having a negative influence on the critical temperature of the final $MgB_2$ films.

First, the high temperature of the substrate (above 800°C) during the annealing leads to a degradation of the SiC diffusion barrier. This effect is shown in Fig.5. Fig.5a presents the R-T dependence of the $MgB_2$ film prepared on top of the SiC/Si substrate (100 nm thick SiC layer) annealed at a temperature of 840°C for 30min prior to the preparation of the $MgB_2$ film. Fig.5b depicts a much improved R-T dependence of the $MgB_2$ film prepared by the same process as in the case of the not pre-annealed substrate. We found the optimal thickness of the SiC layer to be ~ 400 nm when even the pre-annealed substrate has no impact at least on $T_{C0}$ of the $MgB_2$ film.

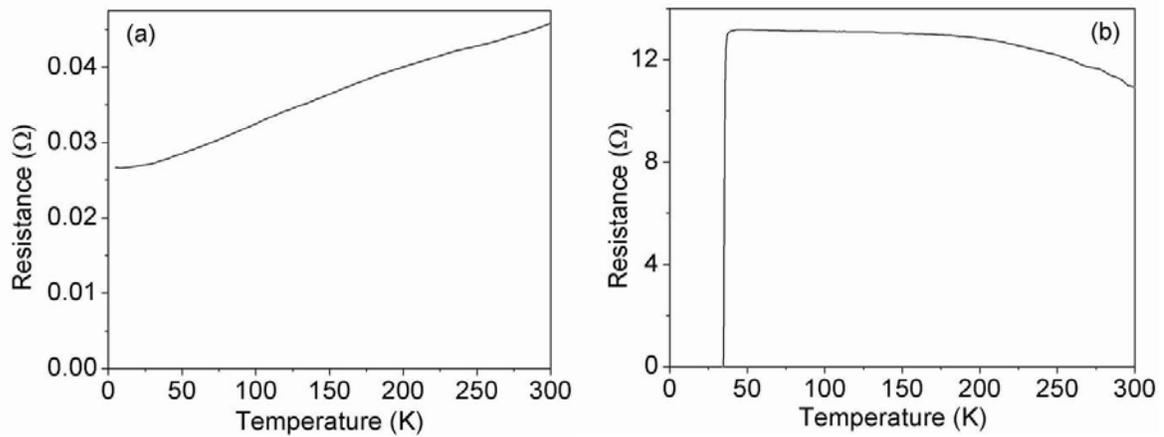

*Fig.5. The dependence of the resistance on the temperature of $MgB_2$ films prepared on top of a pre-annealed (a) and on an as-deposited (b)100 nm thick SiC buffer layer.*

Secondly, we suppose an interaction (reaction) at the interface between the $MgB_2$ and the SiC layer, too.

He et al. [13] reported a reaction between $MgB_2$ and SiC when SiC is mixed with elemental Mg and B in a pressed pellet and annealed at 800°C in a sealed Ta tube. SiC suppressed $T_C$ in that case.

Matsumoto et al. [14] studied the effect of SiC doping of the powder on tube processed $MgB_2$ tapes (SiC doping with a grain size of 30 nm, doping content 1 to 20%). Their results indicated that the doped SiC reacted with Mg forming $Mg_2Si$ at 600°C. The $T_C$ of SiC doped tape is lowered by 3 K compared with the non-doped tape. However, the $J_C$ values of the in-situ processed tapes were much improved by SiC doping.

Zeng et al. in their HPCVD process exposed a single-crystalline SiC substrate to Mg vapor at 720-760°C before $B_2H_6$ was introduced. On the base of the thermodynamic calculation and the thermodynamic $P_{Mg}$-T phase diagram of the SiC-Mg system [8], it was shown that no chemical reaction occurs when the Mg partial pressure is about or below 95% of its vapor pressure at the corresponding temperature in the phase diagram.

In our case (see above), $Mg_2Si$ was observed by a TEM analysis at the surface of the $MgB_2$ film. The formation of $Mg_2Si$ is possible, if it is assumed that Si or SiC diffuse through the whole film up to the surface. However, no $Mg_2Si$ was observed in the bulk of the film. This can be explained if it is assumed that during annealing, the partial pressure of Mg at the B-Mg interface is sufficient for the formation of $Mg_2Si$ (>95% of the vapor pressure at the corresponding temperature in the phase diagram in [8]). In contrast, the partial pressure of Mg decreases in direction to the bulk of the boron film and film-buffer layer interface, hence, no reaction of Mg with SiC is observed (we suppose the diffusion of Mg into the boron film and the volatility of Mg can assist to automatic composition adjustment as a result of the self-limiting adsorption of Mg [8]). We must admit that the RBS analysis was not able to determine the distribution of C and Si in the film. A further analysis was necessary. We characterized our structure by Heavy Ion Elastic Recoil Detection Analysis (ERDA) [15,16]. The ERDA measurements were performed at the Ion Beam Laboratory of the Hahn-Meitner-

Institut using a beam of 350 MeV Au ions. Fig.6 shows the concentration profiles of the $MgB_2/SiC/Si$ structure in the case when no excess metallic Mg is present at the smooth surface (RMS = 9.41nm for 10 $\mu m^2$ area) of the final uniform $MgB_2$ film.

The concentration profiles confirm clearly that Mg and B are not a bilayer anymore and they are diffused into each other. Beside this, mixing of SiC with the $MgB_2$ film and with the Si substrate is evident. These facts support our above suggested assumptions. The presence of oxygen not only at the surface of the film can be linked with amorphous compounds (MgO?) containing oxygen.

Hydrogen and nitrogen in the film are probably a consequence of the addition of $NH_3$ during the preparation of the SiC films.

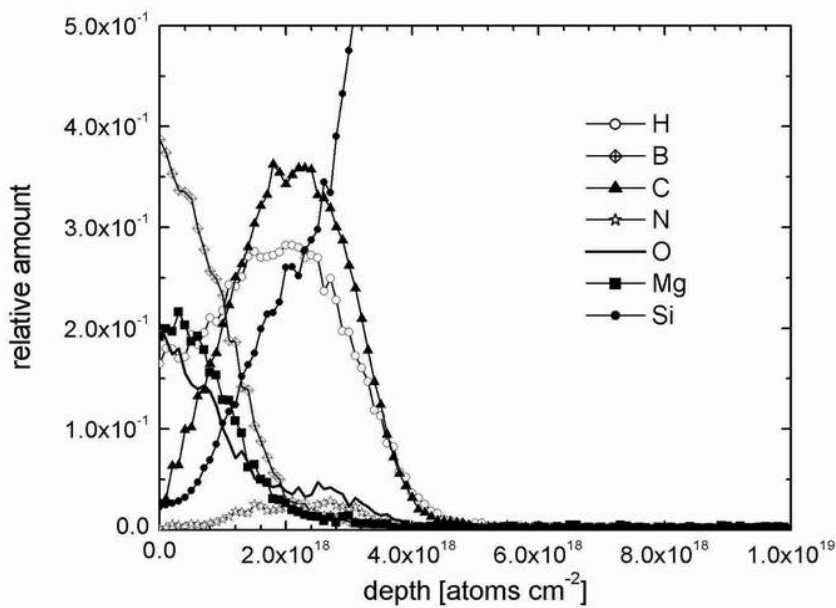

*Fig.6. Concentration profiles of the $MgB_2/SiC/Si$ structure obtained by ERDA.*

## 4. Conclusion

In this paper we have reported on the preparation of the $MgB_2$ thin films by sequential deposition using in-situ annealing on SiC buffered Si substrates. The maximum zero resistance critical temperature $T_{C0}$ = 37.4 K is higher than the previously reported values for the $MgB_2$ films on Si substrates, while the critical current density $J_C$ is comparable with that of polycrystalline films. The presence of the SiC buffer layer enables an high annealing temperature of 830ºC, necessary for the growth process. The TEM analysis revealed the presence of $Mg_2Si$ at the surface of the film, while the bulk of the film is free of this compound, at least in the crystalline form. ERDA confirms the diffusion of SiC into the $MgB_2$ film.

## 5. Acknowledgement

This work was supported by a Slovak Grant Agency for Science (grant No 2/5130/25 and 2/3116/23).